\documentclass[12pt]{article}
\usepackage{epsf}

\newcommand{\be}{\begin{equation}}
\newcommand{\ee}{\end{equation}}
\newcommand{\bmt}{\begin{array}}
\newcommand{\emt}{\end{array}}

\begin{document}

\title{Behavior of the thermal gluon self-energy in the
  Coulomb gauge}
\author{F. T. Brandt$^\dagger$, Ashok Das$^\ddagger$ and J. Frenkel$^\dagger$ 
\\ \\
$^\dagger$Instituto de F\'{\i}sica,
Universidade de S\~ao Paulo\\
S\~ao Paulo, SP 05315-970, BRAZIL\\
$^\ddagger$Department of Physics and Astronomy\\
University of Rochester\\
Rochester, NY 14627-0171, USA}
\maketitle

\bigskip
\noindent

\begin{abstract}
We study, to one loop order, the behavior of the gluon self-energy in
the non covariant Coulomb gauge at finite temperature. The cancellation
of the peculiar energy divergences, which arise in such a gauge, is
explicitly verified in the complete two point function of the
Yang-Mills theory. At high temperatures, the leading $T^2$ term is
determined to be transverse and nonlocal, in agreement with
the results obtained in covariant gauges. The coefficient of the
sub-leading  $\ln(T)$ contribution,  is non transverse but local and
coincides (up to a multiplicative constant) with that of the
ultraviolet pole term of the zero temperature amplitude.
\end{abstract}
In thermal field theory, one is often interested in the contributions
which arise from the region where the loop momenta are of the same
order as the temperature $T$, with $T$  much larger than all the
masses and external momenta 
\cite{weldon:1982aq,kajantie:1985xx,pisarski:1988wc,braaten:1992gm,frenkel:1990br}.
Such {\em hard thermal loop} contributions determine the leading
gauge invariant  terms of the amplitudes at high temperature,
which are important  in {\em resumming} the
QCD thermal perturbation theory  \cite{braaten:1990mz}. 
In general, the coefficients of these leading order terms are not 
directly related to the ultraviolet singular terms of the zero 
temperature amplitudes. In thermal QCD,
for example, the $n$-gluon amplitudes at one-loop order behave
like $T^2$ for high $T$, even though these amplitudes are ultraviolet
finite, at zero temperature, for $n>4$. The hard thermal loop region
is also  relevant for determining the sub-leading, $\ln(T)$ behavior of
the amplitudes.
It has been argued that, in contrast to the behavior of the leading $T^2$
terms, the coefficients of the $\ln(T)$ terms are simply related to
those of the ultraviolet pole terms of the  zero temperature
amplitudes \cite{brandt:1995mc}.

These properties of the amplitudes at high temperature, which have
been verified in covariant and axial gauges, were derived under
the assumption that the integration over the loop energy
$q_0$ is well defined. On the other hand, it is well known that, in
the  Coulomb 
gauge, there are spurious poles at $\vec{q}=0$, leading to divergent
energy  integrals simply because the denominators of some of the
integrands may be independent of $q_0$. 
(For a discussion of this and other related aspects of the
Coulomb gauge  see, for example,  
\cite{mohapatra:1971ue,jackiw:1978ng,christ:1980ku,cheng:1987hu,doust:1987yd,weldon99}).
Consequently, it is not clear, {\em a priori}, whether all the
properties of the thermal amplitudes, derived in covariant and
axial gauges, would continue to hold in the Coulomb gauge as well 
because  of the {\em ill defined} energy integrals.
At zero temperature, there is a proposal to regularize these
singularities using a variant \cite{leibbrandt:1996tn}
of the conventional dimensional regularization \cite{'thooft:1972fi}.
However, it has also been pointed out that, even though individual
Feynman diagrams can have divergent energy integrals in the Coulomb gauge,
such divergences might cancel when all the contributions to a given amplitude
are added together. This has been checked,
using the conventional dimensional regularization, in
the case of the one-loop self-energy for the gluon at zero temperature
\cite{zwanziger:1998ez}.

In this Brief Report we verify explicitly, in one-loop QCD,
that the cancellation of these {\em ill defined} terms takes place at
finite temperature as well. As a  consequence of this, we show 
that all the properties of the hard thermal loop amplitudes, alluded to
earlier, continue to hold even in the Coulomb gauge. Thus, we show
that the  leading $T^2$ term in the gluon self-energy is nonlocal and
is gauge invariant (namely, it is transverse and has the same value
as  in other gauges). The $\ln(T)$
term, on the other hand, is local but non transverse, with the
coefficient coinciding (up to a factor) with that of the ultraviolet
pole term of  the zero temperature amplitude in the Coulomb
gauge. This latter property allows us to determine directly, from the
self-energy for the Coulomb field ($00$ component), the $\ln(T)$
correction to the
effective coupling constant at high temperature. This
simple behavior arises essentially because the Coulomb field is decoupled from
the ghosts  \cite{frenkel:1976bi}.

To carry out the computation at finite temperature, we use the
analytically  continued imaginary time formalism \cite{lebellac:book96},
where the integration over the loop energy is replaced by a
summation over the discrete values $q_0=2\pi\, i\, l\, T$, where $l$ is an
integer. The diagrams which contribute to the gluon self-energy, at
one-loop, are shown in Fig. 1.
\begin{figure}
    \epsfbox{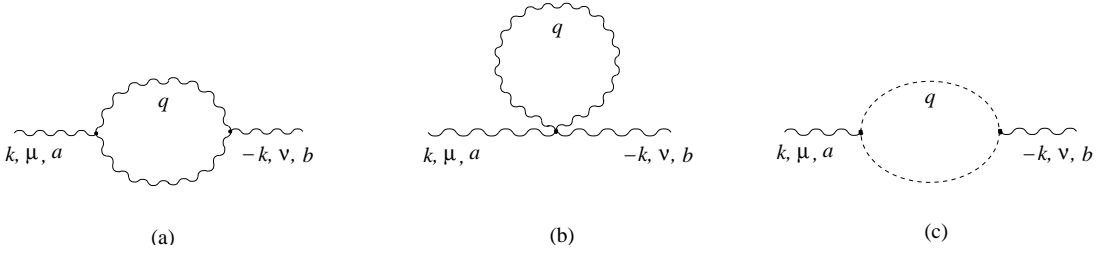}

\caption{One-loop diagrams which contribute to the gluon-self energy. 
Wavy and dashed lines denote respectively gluons and ghosts.} \label{fig1}
  \end{figure}

In the Coulomb gauge, the gluon propagator can be written as
\be\label{1}
D_{\mu\nu}^{ab}(q)=\frac{\delta^{ab}}{q^2}\left[\eta_{\mu\nu} + 
\frac{n^2\, q_\mu q_\nu - q\cdot n\left(q_\mu n_\nu + q_\nu n_\mu\right)}
{\vec q^2}\right],
\ee
where $n^\mu=(1,0,0,0)$, while the ghost propagator has the form
\be\label{2}
D^{ab}(q)=\frac{\delta^{ab}}{\vec q^2}.
\ee
This propagator is independent of energy and this is one of the
reasons that divergent energy integrals arise.

Let us consider first the ghost loop contribution to the gluon
self-energy shown in Fig. (1c). Using the
appropriate ghost-gluon-ghost vertex  in the Coulomb gauge, one
finds that, at finite temperature, this
graph leads to the contribution
\begin{eqnarray}\label{3}
\Pi^{ij,ab}_{(ghost)}=\frac{\delta^{ab} N g^2\, T}{2}\sum_{q_0}\int
\frac{{\rm d}^d\vec q}{(2\pi)^d}
\frac{\left(2 q^i q^j + k^i q^j + k^j q^i\right)}
{\vec q^2\left(\vec q + \vec k\right)^2}\nonumber\\
= \frac{\delta^{ab} N g^2\, \pi^{\frac{d}{2}}}{2\,(2\pi)^d}
{\Gamma(1-\frac d 2)}
\frac{\Gamma^2(\frac d 2)}{\Gamma(d)}
|\vec k|^{d - 4}\,T
\sum_{q_0}\left[\delta^{ij}\vec k^2 + (d-2) k^i k^j\right]
\end{eqnarray}
Here  $g$ is the gauge coupling constant, $N$ the color factor of
SU($N$), $d$ the space dimension ($=3$ in four space-time
dimensions) and $\Gamma$ denotes the gamma function
\cite{gradshteyn}. The  divergence associated with the $q_0$ sum, in
Eq. (\ref{3}),  is now obvious. However, it is easy to check that all
such terms cancel out when we take  into
consideration similar contributions which arise from the $00$
components of the internal gluon propagators in the other diagrams
in Fig. 1. In fact, adding all such contributions, we obtain
\be\label{4}
\Pi^{ij,ab}_{(00)}+\Pi^{ij,ab}_{(ghost)}=\frac{\delta^{ab} N g^2\,
T}{2}\sum_{q_0}\int \frac{{\rm d}^d \vec q}{(2\pi)^d}
\frac{\left(k^i k^j+ k^i q^j + k^j q^i \right)}
{\vec q^2\left(\vec q + \vec k\right)^2}=0
\ee
which follows directly from using the standard dimensional
regularization. Note that the
cancellation between the ghost contribution and those from the unphysical
components of the gluon field, occurs in any space dimension provided
one  uses systematically
conventional dimensional regularization \cite{'thooft:1972fi}.

Since the {\em ill defined} terms cancel, we can now proceed with the
standard  method for evaluating the remaining finite temperature
contributions, which is facilitated by the use of the relation
\cite{lebellac:book96}
\begin{eqnarray}\label{5}
T\sum_{l=-\infty}^{\infty} I(q_0=\pi\, i\, l\, T) =
\frac{1}{4\pi\, i}\int_{-i\infty}^{+i\infty}{{\rm d}q_0}
\left[I(q_0) + I(-q_0)\right] + \nonumber \\
\frac{1}{2\pi\, i}\int_{-i\infty+\delta}^{+i\infty+\delta}{\rm d}q_0
\left[I(q_0) + I(-q_0)\right]\frac{1}{\exp\left(q_0/T\right) -1}
\end{eqnarray}
Here $I(q_0)$ is given by an integral over the space components $\vec
q$ and $\delta\rightarrow 0^+$.
The first term on the right-hand side represents the
zero-temperature part of the amplitude while the second term contains the
thermal corrections which involve the Bose-Einstein distribution function. In
the thermal part, the contour in the $q_0$ complex plane may now be closed in 
the right half-plane and the $q_0$ integration
performed by evaluating the contributions from the poles
of the gluon propagator. In this way, the second term in Eq. (\ref{5})
may be expressed in terms of forward scattering amplitudes of
on-shell thermal particles with four momentum $q^\mu =(|\vec q|,\vec q)$, 
as illustrated in  Fig. 2  \cite{frenkel:1990br,brandt:1997se}.
\begin{figure}
    \epsfbox{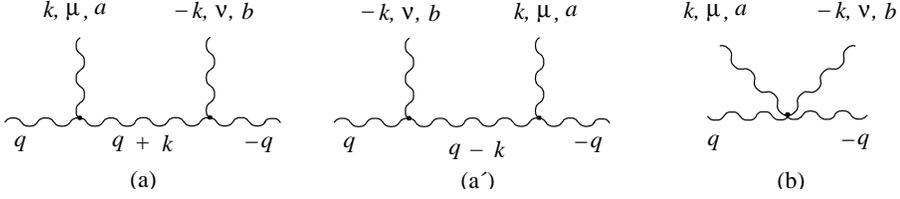}
\caption{Forward scattering graphs associated with diagrams (1a) and
  (1b).}
\label{fig2}
  \end{figure}

Denoting by $A^{\mu\nu,ab}(q,k)$ the total forward scattering amplitude,
where the sum over the polarizations and the color states of the thermal
gluon is to be understood, we can write the thermal contributions in
terms of a momentum integral of $A^{\mu\nu,ab}$ as follows
\be\label{6}
\Pi^{\mu\nu,ab}_{thermal}=
\int \frac{{\rm d}^3 \vec q}{(2\pi)^3}\frac{1}{2|\vec q|}
\frac{1}{\exp{(|\vec q|/T)}-1}
\left[A^{\mu\nu,ab}(q,k) + A^{\mu\nu,ab}(-q,k)\right]_{q_0=|\vec q|}
\ee
where we have set the space dimension to $d=3$, which is the case of
interest to us. Note that the temperature
provides a natural ultraviolet cut-off for the thermal corrections. We
may now extract from Eq. (\ref{6}) a series of high-temperature
contributions which arise from the region of large $q$. In the
hard thermal region, we can use the expansion
\be\label{7}
\left.\frac{1}{\left(q+k\right)^2}\right|_{q_0=|\vec q|} =
\frac{1}{2 q\cdot k} - \frac{k^2}{\left(2 q\cdot k\right)^2} +
\frac{k^4}{\left(2 q\cdot k\right)^3} + \cdots
\ee
in the denominators of the forward scattering amplitude, and expand
its numerator similarly in powers of $k/q$. One thus gets 
for $A^{\mu\nu,ab}$ in Eq. (\ref{6}) terms which are homogeneous in 
$q$ of degree $1,0,-1,-2$ and so on. 
The first term has a denominator of the form $1/(q\cdot k)$ and a
numerator which is quadratic in $q$ and independent of $k$. Such odd terms
cancel out in Eq. (\ref{6}) by symmetry under $q\rightarrow -q$. The
next contributions are down by a power of $k/q$ and arise from the
terms in $A^{\mu\nu,ab}$ which are of zero degree in $q$. Such terms
yield the leading $T^2$ contributions. The next non-vanishing
contributions come from those terms in $A^{\mu\nu,ab}$ which are of
degree $-2$ in $q$. By power counting, these give rise to the
$\ln(T)$ contributions.
Performing the integration over $q$, we determine the leading $T^2$
contribution to be
\be\label{8}
\Pi^{\mu\nu,ab}_{(T^2)}=\frac{g^2 N\delta^{ab}}{48\pi}\,T^2
\int{\rm d}\Omega\left(
\frac{\hat q^\mu k^\nu + \hat q^\nu k^\mu}{\hat q\cdot k} 
-\frac{\hat q^\mu \hat q^\nu k^2}{(\hat q\cdot k)^2} -\eta^{\mu\nu}
\right)
\ee
where $\int{\rm d}\Omega$ denotes the integration over the directions
of $\hat q=\vec q/|\vec q|$ and $\hat q^\mu=(1,\hat q)$.
This contribution, which is clearly non local and manifestly
transverse, agrees with the well
known hard thermal loop result obtained in 
covariant and axial gauges  \cite{weldon:1982aq,frenkel:1990br}.
Such a gauge independent contribution has a physical interpretation
in terms of plasma frequencies and screening lengths \cite{lebellac:book96}.

Furthermore, we have determined the $\ln(T)$ terms in
the  thermal part of the gluon self-energy to be
\be\label{9}
\Pi^{\mu\nu,ab}_{(\ln T)}=\frac{g^2 N\delta^{ab}}{8\pi^2}
\left[
k^\mu k^\nu - k^2 \eta^{\mu\nu} + \frac 4 3 
\frac{k\cdot n}{n^2}
\left(k^\mu n^\nu + k^\nu n^\mu\right) - \frac 8 3
\frac{k^2}{n^2}
n^\mu n^\nu \right]\ln\left(\frac{T}{\kappa}\right)
\ee
where $n^\mu=(1,0,0,0)$ and
$\kappa$ is a typical external momentum scale. This expression which is 
local, but non transverse, coincides with the
coefficient (up to a factor) of the ultraviolet pole term of the
zero-temperature self-energy in the Coulomb gauge
\cite{leibbrandt:1996tn}, as expected \cite{brandt:1995mc}.

The above correspondence implies that the coefficient of $\ln(T/\kappa)$
in Eq. (\ref{9})
must be the same as the coefficient of  $\ln(\kappa/\mu)$ in the
renormalized amplitude at zero temperature, where $\mu$ is the renormalization
scale. This property allows us to determine, in a simple way, the $\ln(T)$
corrections in the running coupling constant at high temperature.
To this end, we use the fact 
that the logarithmic contribution to the running coupling
constant $\bar g(\kappa/\mu)$ at $T=0$, can be determined directly
from the renormalized Coulomb field amplitude  \cite{frenkel:1976bi}
\be\label{10}
\Pi^{00,ab}_{(\ln \mu)}=
\delta^{ab}\vec k^2 \left[\frac{11\,Ng^2}{24\pi^2}\,
\ln\left(\frac{\kappa}{\mu}\right)\right]
\ee
From the temperature dependent part of the $00$ amplitude in Eq. (\ref{9}),
we see that the $\kappa$ dependence cancels in the total amplitude
so that the complete Coulomb thermal amplitude contains only a logarithmic
factor $\ln(T/\mu)$. This term will then determine the logarithmic
contribution to the running coupling constant $\bar g(T/\mu)$ at high
temperature.

\bigskip

\noindent
We would like to thank Professor J. C. Taylor for a helpful correspondence.
This work was supported in part by U.S. Dept. Energy Grant DE-FG
02-91ER40685, NSF-INT-9602559 as well as by CNPq, Brazil.


\newpage

\begin{thebibliography}{10}
\bibitem{weldon:1982aq}
H.~A. Weldon, Phys. Rev. {\bf D26},  1394  (1982).
\bibitem{kajantie:1985xx}
K. Kajantie and J. Kapusta, Ann. Phys. {\bf 160},  477  (1985).
\bibitem{pisarski:1988wc}
R.~D. Pisarski, Nucl. Phys. {\bf B309},  476  (1988).
\bibitem{braaten:1992gm}
E. Braaten and R.~D. Pisarski, Phys. Rev. {\bf D45},  1827  (1992).
\bibitem{frenkel:1990br}
J. Frenkel and J.~C. Taylor, Nucl. Phys. {\bf B334},  199  (1990);
%
{\it ibid} Nucl. Phys. {\bf B374},  156  (1992).
\bibitem{braaten:1990mz}
E. Braaten and R.~D. Pisarski, Nucl. Phys. {\bf B337},  569  (1990);
%
{\it ibid} Nucl. Phys. {\bf B339},  310  (1990).
\bibitem{brandt:1995mc}
F.~T. Brandt and J. Frenkel, Phys. Rev. Lett. {\bf 74},  1705  (1995);\\
Phys. Rev. {\bf D55},  7808  (1997).\\
F.~T. Brandt, J. Frenkel, and F.~R. Machado, Phys. Rev. {\bf D61},  125014
  (2000).
\bibitem{mohapatra:1971ue}
R.~N. Mohapatra, Phys. Rev. {\bf D4},  378  (1971).
\bibitem{jackiw:1978ng}
R. Jackiw, I. Muzinich, and C. Rebbi, Phys. Rev. {\bf D17},  1576  (1978).
\bibitem{christ:1980ku}
N.~H. Christ and T.~D. Lee, Phys. Rev. {\bf D22},  939  (1980).
\bibitem{cheng:1987hu}
H. Cheng and E.-C. Tsai, Phys. Rev. {\bf D34},  3858  (1986);
{\it ibid} {\bf D36},  3196  (1987).
\bibitem{doust:1987yd}
P.~J. Doust and J.~C. Taylor, Phys. Lett. {\bf B197},  232  (1987);\\
J.~C. Taylor, in {\it Physical and nonstandard gauges}, Lecture Notes
on Physics, Vol. 361, ed. P. Gaigg, W. Kummer, M. Schweda, 
Springer-Verlag,
Berlin, 137 (1990).
\bibitem{weldon99}
H. A. Weldon, Annals Phys. {\bf 271}, 141, (1999).
\bibitem{leibbrandt:1996tn}
G. Leibbrandt and J. Williams, Nucl. Phys. {\bf B475},  469  (1996).\\
G. Heinrich and G. Leibbrandt, Nucl. Phys. {\bf B575},  359  (2000).
\bibitem{'thooft:1972fi}
G. 't~Hooft and M. Veltman, Nucl. Phys. {\bf B44},  189  (1972).
\bibitem{zwanziger:1998ez}
D. Zwanziger, Nucl. Phys. {\bf B518},  237  (1998)\\
L. Baulieu and D. Zwanziger, Nucl. Phys. {\bf B548},  527  (1999).
\bibitem{frenkel:1976bi}
J. Frenkel and J.~C. Taylor, Nucl. Phys. {\bf B109},  439  (1976).
\bibitem{lebellac:book96}
J. ~I. Kapusta, {\em Finite Temperature Field Theory} 
(Cambridge University Press, Cambridge, England, 1989).\\
M.~L. Bellac, {\em Thermal Field Theory} (Cambridge University Press,
  Cambridge, England, 1996).\\
A. Das, {\em Finite Temperature Field Theory} (World Scientific,
NY 1997).
\bibitem{gradshteyn}
I.~S. Gradshteyn and M. Ryzhik, {\em Tables of Integral Series and Products}
  (Academic, New York, 1980).
\bibitem{brandt:1997se}
F.~ T. Brandt and J. Frenkel,
Phys. Rev. {\bf D56}, 2453 (1997).\\
F.~ T. Brandt, A. Das, J. Frenkel and A. J. da Silva,
Phys. Rev. {\bf D59}, 065004 (1999).

\end{thebibliography}
\end{document}